\documentclass[final,5p,times,twocolumn]{elsarticle}

\usepackage{graphicx}
\usepackage[dvips]{color}
\usepackage[figuresright]{rotating}

\usepackage{amssymb}
\usepackage{amsmath}

\journal{Nucl. Instr. and Meth. in Phys. Res. A}

\begin{document}  

\begin{frontmatter}


\title{Hadron calorimeter with MAPD readout in the NA61/SHINE experiment}

\author[INR]{A.~Ivashkin}  
\author[JINR]{F.~Akhmadov}  
\author[Geneva]{R.~Asfandiyarov} 
\author[Geneva]{A.~Bravar} 
\author[Geneva]{A.~Blondel} 
\author[Warsaw]{W.~Dominik}  
\author[KFKI]{Z.~Fodor}  
\author[UFK]{M.~Gazdzicki}  
\author[INR]{M.~Golubeva}  
\author[INR]{F.~Guber}
\author[Geneva]{A.~Hasler}
\author[Geneva]{A.~Korzenev}  
\author[Chile]{S.~Kuleshov}  
\author[INR]{A.Kurepin}  
\author[KFKI]{A.~Laszlo}  
\author[INR]{V.~Marin}  
\author[INR]{Yu.~Musienko}  
\author[INR]{O.~Petukhov}  
\author[Bergen]{D.~R\"ohrich}  
\author[INR]{A.~Sadovsky}  
\author[JINR]{Z.~Sadygov}  
\author[KFKI]{T.~Tolyhi}  
\author[Zecotek]{F.~Zerrouk}  

\address[INR]{Institute for Nuclear research RAS, 117312 Moscow, Russia} 
\address[JINR]{Joint Institute for Nuclear research RAS, Dubna, Russia}
\address[Geneva]{University of Geneva, Switzerland} 
\address[Warsaw]{Faculty of Physics, University of Warsaw, Warsaw, Poland} 
\address[KFKI]{KFKI Research Institute for Particle and Nuclear Physics, Budapest, Hungary}
\address[UFK]{Universities of Frankfurt and Kielce}
\address[Chile]{The Universidad Tecnica Federico Santa Maria, Valparaiso, Chile}
\address[Bergen]{University of Bergen, Bergen, Norway}  
\address[Zecotek]{Zecotek Imaging Systems Pte. Ltd., Singapore}

\begin{abstract}  

The modular hadron calorimeter with micro-pixel avalanche photodiodes readout for the NA61/SHINE experiment at the CERN SPS  is presented.  The calorimeter consists of 44 independent modules with lead-scintillator sandwich
 structure. The light from the scintillator tiles  
is captured by and transported  with WLS-fibers  embedded in scintillator grooves.  The construction provides a 
longitudinal  segmentation of the module in 10 sections with
independent MAPD readout. MAPDs with   
pixel density of $~10^{4}$/mm$^2$ ensure good linearity of calorimeter response in a wide dynamical 
range.  The performance of the calorimeter prototype in a beam test is reported.   

\end{abstract}  


\end{frontmatter}


\section{Introduction}  

The goal of the NA61/SHINE experiment at CERN ~\cite{NA61:proposal} is the search for the critical  
end-point and the onset of deconfinement in ion-ion collisions. A two dimensional  
scan of the phase diagram of strongly interacting matter will be done by changing  
the ion beam energy at the SPS  (13-158 AGeV) and the size of the colliding  
systems. The critical point would be indicated by a maximum in the fluctuation of  
the particle multiplicity and other physical observables. The onset of deconfinement  
is revealed by rapid changes in the hadron production properties.  
Study of  fluctuations due to properties of strongly interacting matter  requires   
a very precise  control  over  fluctuations caused by the variation of the number of interacting nucleons.   
The latter result from ''trivial'' event-by-event   changes of the collision geometry.   
The number of interacting nucleons can be determined by measurements of  the number of non-interacting   
nucleons from projectile nuclei (projectile spectators) by  a very forward hadron calorimeter.   
The designed calorimeter for the NA61/SHINE experiment is called  the Projectile Spectator Detector (PSD).   
Basic design requirements are good  energy resolution,   
${\sigma_{E}\over E}<{60\% \over \sqrt{E(GeV)}}$,   
and good transverse uniformity of this resolution. Fully compensating modular lead/scintillator hadron   
calorimeters~\cite{Alekseev:2001,Fujii:2000} meet these requirements.  

\section{Calorimeter construction}  
The PSD calorimeter consists of 44  modules  which cover a transverse area of $~$120x120~cm$^{2}$.   
Each  module consists of 60 lead/scintillator layers with 16~mm and 4~mm thickness,   
respectively. The lead/scintillator plates are tied together with   
0.5~mm thick steel tape and placed in a box made of 0.5~mm thick steel.  Steel tape and box are spot-welded 
together providing appropriate mechanical rigidity.  The full length of modules corresponds to 5.7 nuclear 
interaction lengths. 
To fit the PSD  transverse dimensions to the size of the spectator spots the distance between the NA61 target 
and the calorimeter is increased with increasing collision energy  from 17~m to 23~m. 

The central part of the PSD consists of 16 small   
modules with transverse dimension of 10x10~cm$^{2}$ and weight ~120 kg each. Such fine transverse segmentation  
decreases the spectator occupancy in one module and improves the reconstruction of the reaction plane. The outer part 
of PSD the comprises 28 larger  20x20~cm$^{2}$ modules with a weight of ~500 kg each.   The mechanical rigidity of 
these heavy modules was enhanced by a slight modification of their structure. Namely, one 16 mm lead layer 
in the middle of the module was replaced by a steel plate with similar nuclear interaction length. 

Light read-out is provided by Kyraray  Y11 WLS-fibers embedded in round grooves in the  scintillator plates.   
The WLS-fibers from each 6 consecutive scintillator tiles are collected   
together in a single optical connector at the end of the module.   
Each of the 10 optical connectors at the downstream face of the module  is read-out by a single photo-diode.   
The longitudinal segmentation into 10 sections ensures good uniformity of   
light collection along the module and delivers information on the type of particle  
which caused the observed particle shower.    10 photodetectors per module are placed at the rear side of 
the module together with the  front-end-electronics. 

\section{Choice of photodetectors}  

Longitudinal segmentation of calorimeter modules requires 10 individual photodetectors per module for  
the signal readout. Silicon photomultipliers SiPMs or  micro-pixel avalanche photodiodes, MAPDs~\cite{Sadygov:2006} are an optimum choice due to their  
remarkable properties such as high internal gain, compactness, low cost and immunity to the nuclear  
counter effect. Moreover, calorimetry applications have some specific requirements such 
as large dynamical range and linearity of photodetector response to intense light pulses.
As known, the dynamic range and linearity of MAPDs are limited by the finite number of pixels. Most of the 
existing types of MAPDs with individual surface resistors have a pixel density $~10^3$ pixels/mm$^2$. Such 
a limited number leads to a serious restriction of MAPD application in calorimetry, where the number of 
detected photons is comparable and even larger than the pixel number. The effect of saturation, when a few 
photons hit the same pixel, leads to significant non-linear MAPD response to light pulses with high 
intensity. Evidently, the MAPD has linear response only if the number of pixels is much larger than the number of 
incident photons. This feature represents a disadvantage of MAPDs compared to the traditional PMTs. 
However,this drawback is essentially reduced for MAPDs with individual micro-well 
structure~\cite{Sadygov:2006}, where a pixel density of 10$^4$/mm$^2$ and higher is achievable.  
The above considerations motivated the choice of photodetectors: photodiodes of MAPD-3A type produced by 
Zecotek Photonics Inc. (Singapore)~\cite{site:zecotek} were selected for the readout of the PSD hadron calorimeter. 
These MAPDs have a pixel density of 15000/mm$^2$. Their 3x3 mm$^2$  active area  nicely fits  the size of the 
WLS-fiber bunch from one longitudinal section of a PSD module and provides a total number of pixels of more than 
10$^5$ in one photodetector. 

\begin{figure}[htb]  
  \centering  
     \includegraphics[width=0.45\textwidth]{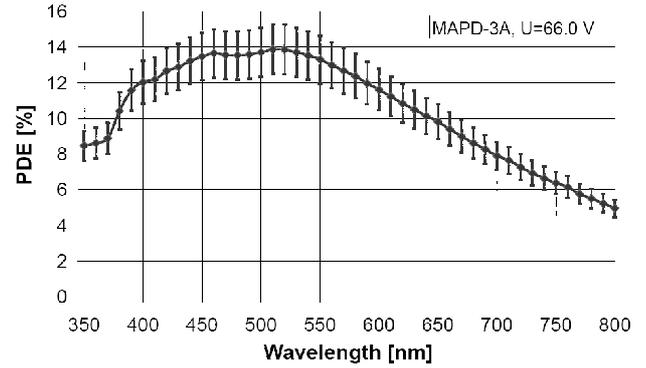}  
\caption{  Spectral response of MAPD-3A: dependence of the photon detection efficiency PDE on the wavelength of light. 
The error bars correspond to the systematic uncertainty. }  
\label{figure_1}  
\end{figure}

\begin{figure}[htb]  
  \centering  
     \includegraphics[width=0.45\textwidth]{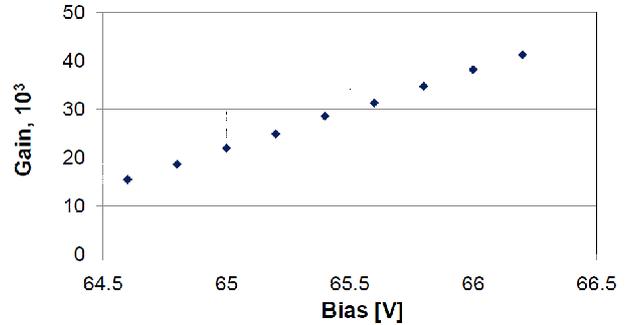}  
\caption{  Dependence of MAPD-3A gain on the bias voltage.  
}  
\label{figure_2}  
\end{figure}

The MAPD-3A photon detection efficiency (PDE) for Y11 WLS-fiber emission spectrum at 510 nm reaches 15\%   
and is rather similar to PMT performance, see Fig.~\ref{figure_1}. The operation voltage for different samples of 
MAPD-3A ranges from 65 V to 68 V. The maximum achieved gain is  5$\times$10$^4$, see Fig.~\ref{figure_2}.

\begin{figure}[htb]  
  \centering  
  \includegraphics[width=0.45\textwidth]{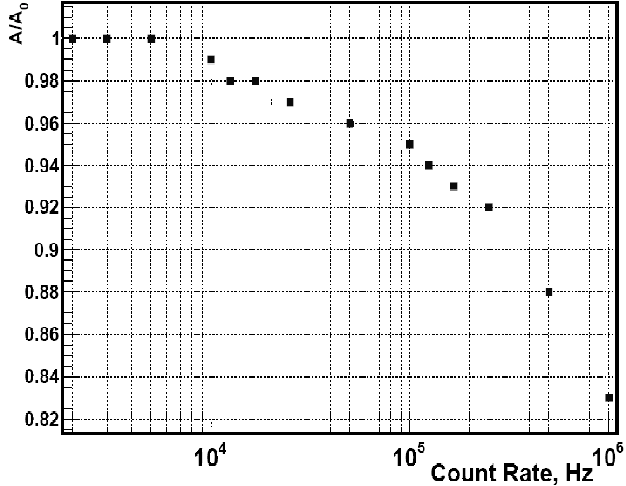}  
\caption{    Dependence of normalized MAPD-3A amplitude on the frequency of light pulses with fixed intensity. 
The MAPD-3A amplitude  at the frequency 1 kHz corresponds to 1500 photoelectrons.  
}  
\label{figure_3}  
\end{figure}

Since the PSD calorimeter  has no beam hole to ensure maximum acceptance for the spectators,  the central part 
of the PSD will be irradiated by the ion beam with an intensity up to 10$^5$ Hz. Therefore, one of the stringent 
requirements for the calorimeter readout is high count rate capability, at least in the central region. Here, 
the average amplitude in one longitudinal section of a PSD module is expected to be about 1500 photoelectrons. 
In other words,  the recovery time of the selected MAPD-3A photodiode must be fast enough to ensure stable 
amplitudes at signal frequencies up to 100 kHz. To check the count rate capability of the MAPD-3A the dependence 
of its amplitude on the frequency of light pulses was measured. The stability of the  amplitude of the pulses at different 
frequencies from light emitting diode was checked by a normal PMT. The obtained behavior is presented in
Fig.~\ref{figure_3}. As seen, the MAPD-3A amplitude would drop about 5\%  for the maximum beam intensity
 foreseen in the NA61/SHINE experiment . 

\begin{figure}[htb]  
  \centering  
 \includegraphics[width=0.45\textwidth]{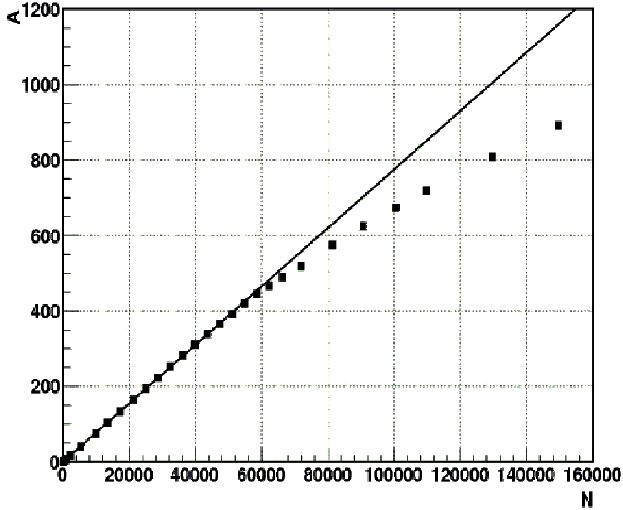}  
\caption{    The dependence of MAPD amplitude (in relative units) on the number of incident photons. 
}  
\label{figure_4}  
\end{figure}

To check the photodetector linearity measurements of MAPD amplitudes were performed with light pulses of different intensity. 
The number of incident photons was determined by a reference photodiode with known quantum efficiency and gain equal to one. 
The dependence of the MAPD amplitude on the number of incident photons is shown in Fig.~\ref{figure_4}.
As seen, MAPD linearity is preserved for light pulses with a number of photons up to 6$\times$10$^4$.  
Taking into account the photon detection efficiency of  about 25\% for the tested MAPD (with higher photon detection efficiency, 
but with the same pixel density as the MAPD-3A type) linear response to amplitudes up to 15000 photoelectrons is expected. 

The reported comprehensive studies confirm that the selected MAPD-3A photodetectors satisfy the requirements of 
the NA61/SHINE experiment. At present, 320 MAPD-3A samples are installed in the available PSD modules and show stable 
operation during two months of calibration and beam technical  beam runs.

\section{ Beam tests of the PSD calorimeter}  

\begin{figure}[htb]  
  \centering  
     \includegraphics[width=0.45\textwidth]{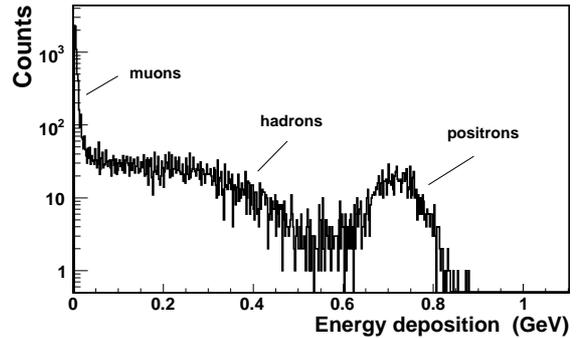}  
\caption{    Energy spectrum measured in the first section of the  
central module for a 30 GeV hadron beam containing a fraction of muons and positrons.  
}  
\label{figure_5}  
\end{figure}

At present, the major part (16 small and 16 large modules)  of the PSD calorimeter is assembled   
and installed in the  NA61 experimental area. The  construction of the full detector   
will be completed during the coming months.  To check the performance of the calorimeter several tests   
using hadron beams at various energies were performed.   
For this purpose a PSD module array of nine small modules  (3$\times$3 array) was assembled.   
The first test at high energies was done in the NA61 H2  
beam-line at the CERN SPS. The second test with low energy beams was carried out   
in the T10 beam-line at the CERN PS. The hadron beam energy ranged from 20~GeV to 158~GeV during 
the SPS test, and from 1~GeV to 6~GeV during the PS test. In both tests the calibration   
of all read-out channels was preformed  with a muon beam.   
In order to derive  the full set of 90 calibration coefficients a muon beam scan was performed   
for all 9 modules  and for 10 longitudinal sections in each module.   
The energy resolution of   the central module of the tested array was estimated  from data taken with  
hadron beams.  Note, that the  SPS hadron beam at low energies contains a significant fraction of muons 
and positrons.  The spectrum of deposited energy  in the first section of the central module exposed to the  30~GeV beam is shown  in Fig.~\ref{figure_5}.   
The right-side peak in the spectrum corresponds to full positron  energy absorption in the first longitudinal section   
that might be regarded as an electromagnetic  calorimeter with rough sampling.   
The energy resolution for positrons at 30~GeV is about 6.5\%.  

\begin{figure}[htb]  
  \centering  
     \includegraphics[width=0.45\textwidth]{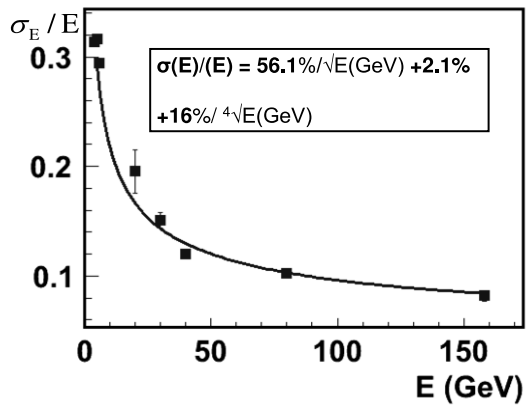}  
\caption{    The energy resolution of the tested PSD prototype as a function of  
hadron beam energy . The solid line is the fit of experimental data by the function   
shown in the insert.  
}  
\label{figure_6}  
\end{figure}

The dependence of measured energy resolution on the pion energy is shown  in Fig.~\ref{figure_6}.   
The tested prototype with 30$\times$30~cm$^{2}$   transverse size is too small to contain the entire   
hadron shower. Therefore, a non-negligible lateral shower leakage is expected.   
Monte Carlo  simulations confirm that about 16\% of hadron shower energy escapes from the tested  
array. The influence of shower leakage on the energy resolution   
was considered in~\cite{Acosta:1991, Badier:1994}, where a third term in addition to the   
stochastic and constant ones was added in the parameterization of the resolution.   
The fit of the experimental data with the three-term formula   
(Fig.~\ref{figure_6}) gives the coefficient of the stochastic term   
equal to 56.1\% and of the constant term equal to 2.1\% at the fixed leakage term of 16\%.   
A non-zero constant term might   
indicate that the selected lead/scintillator sampling does not provide full compensation.   
In order  
to reduce the lateral shower leakage a beam test of   
the calorimeter with a larger size will be performed soon.  
 
The authors thank all members of the NA61/SHINE collaboration for fruitful discussions and   
permanent attention. This work was supported by the Polish Ministry of   
Science and Higher Education (grant PBP 2878/B/H03/2010/38), the Russian Foundation for Basic Research 
(grant 09-02-00664),  Cooperation Programme Switzerland-Russia (grant STCP-RUSSIA/S17604)
and by FONDECYT (Chile) under the project 1100582  
and Centro-Cient\'\i fico-Tecnol\'{o}gico de Valpara\'\i so PBCT ACT-028.

\end{document}